# Magnetic field gradient driven dynamics of isolated skyrmions and antiskyrmions in frustrated magnets

J. J. Liang[1], J. H. Yu[1], J. Chen[1], M. H. Qin[1,*], M. Zeng[1], X. B. Lu[1], X. S. Gao[1], and J. –M. Liu[2,†]

[1]*Institute for Advanced Materials, South China Academy of Advanced Optoelectronics and Guangdong Provincial Key Laboratory of Quantum Engineering and Quantum Materials, South China Normal University, Guangzhou 510006, China*

[2]*Laboratory of Solid State Microstructures and Innovative Center for Advanced Microstructures, Nanjing University, Nanjing 210093, China*

**[Abstract]** The study of skyrmion/antiskyrmion motion in magnetic materials is very important in particular for the spintronics applications. In this work, we study the dynamics of isolated skyrmions and antiskyrmions in frustrated magnets driven by magnetic field gradient, using the Landau-Lifshitz-Gilbert simulations on the frustrated classical Heisenberg model on the triangular lattice. A Hall-like motion induced by the gradient is revealed in bulk system, similar to that in the well-studied chiral magnets. More interestingly, our work suggests that the lateral confinement in nano-stripes of the frustrated system can completely suppress the Hall motion and significantly speed up the motion along the gradient direction. The simulated results are well explained by the Thiele theory. It is demonstrated that the acceleration of the motion is mainly determined by the Gilbert damping constant, which provides useful information for finding potential materials for skyrmion-based spintronics.

Keywords: skyrmion dynamics, field gradient, frustrated magnets

PACS numbers: 12.39.Dc, 66.30.Lw.Kw, 75.10.Jm

Email: [*]qinmh@scnu.edu.cn, [†]liujm@nju.edu.cn

## I. INTRODUCTION

Magnetic skyrmions which are topological defects with vortex-like spin structures have attracted extensive attention since their discovery in chiral magnets due to their interesting physics and potential applications in spintronic devices.[1-4] Specifically, the interesting characters of skyrmions such as the topological protection[5], the ultralow critical currents required to drive skyrmions (~$10^5$ Am$^{-2}$, several orders of smaller than that for domain-wall manipulation)[3,6], and their nanoscale size makes them proposed to be promising candidates for low power consumption magnetic memories and high-density data processing devices. Theoretically, the cooperation of the energy competition among the ferromagnetic, Dzyaloshinskii-Moriya (DM), and the Zeeman couplings and the thermal fluctuations is suggested to stabilize the skyrmions.[7,8] Moreover, the significant effects of the uniaxial stress on the stabilization of the skyrmion lattice have been revealed in earlier works.[9-12] On the skyrmion dynamics, it has been suggested that the skyrmions in chiral magnets can be effectively modulated by spin-polarized current,[13-16] microwave fields,[17] magnetic field gradients,[18,19] electric field gradients,[20,21] temperature gradients[22] etc. So far, some of these manipulations have been realized in experiments.[23]

Definitely, finding new magnetic systems with skyrmions is essential both in application potential and in basic physical research.[24] More recently, frustrated magnets have been suggested theoretically to host skyrmion lattice phase. For example, skyrmion crystals and isolated skyrmions have been reported in the frustrated Heisenberg model on the triangular lattice.[25,26] In this system, it is suggested that the skyrmion crystals are stabilized by the competing ferromagnetic nearest-neighbor (NN) and antiferromagnetic next-nearest-neighbor (NNN) interactions and thermal fluctuations at finite temperatures ($T$) under applied magnetic field $h$. Furthermore, the uniaxial anisotropy strongly affects the spin orders in triangular antiferromagnets and stabilizes the isolated skyrmions even at zero $T$.[26,27]

Compared with the skyrmions in chiral magnets, those in frustrated magnets hold two additional merits. On the one hand, the skyrmion lattice constant is typically an order of magnitude smaller than that of chiral magnets, and higher-density data processing devices are expected. On the other hand, the skyrmions are with two additional degrees-of-freedom (vorticity and helicity) due to the fact that the exchange interactions are insensitive to the direction of spin rotation. As a result, both skyrmion and antiskyrmion lattices are possible in frustrated magnets which keep the $Z_2$ mirror symmetry in the *xy* spin component. Furthermore, the dynamics of skyrmions/antiskyrmions is probably different from that of chiral magnets, as revealed in earlier work which studied the current-induced dynamics in nanostripes of frustrated magnets.[28] It has been demonstrated that the spin states formed at the edges create multiple edge channels and guide the skyrmion/antiskyrmion motion.

It is noted that spin-polarized current may not drive the skyrmion well for insulating materials, and other control parameters such as field gradient are preferred. In chiral magnets, for example, the gradient can induce a Hall-like motion of skyrmions, i. e., the main velocity $v_\perp$ (perpendicular to the gradient direction) is induced by the gradient, and a low velocity $v_\parallel$ (parallel to the gradient direction) is induced by the damping effect. Thus, the gradient-driven motion of skyrmions and antiskyrmions in frustrated systems is also expected. Furthermore, it has been suggested that the confined geometry suppresses the current-induced Hall motion of skyrmions and speeds up the motion along the current direction, which is instructive for future applications.[29] In some extent, the gradient-driven motion could also be strongly affected by confining potential in narrow constricted geometries. Thus, as a first step, the field-gradient-induced dynamics of skyrmions and antiskyrmions in bulk frustrated magnets as well as in constricted geometries urgently deserves to be revealed theoretically. However, few works on this subject have been reported, as far as we know.

In this work, we study the skyrmion/antiskyrmion dynamics in frustrated magnets induced by magnetic field gradients using Landau-Lifshitz-Gilbert (LLG) simulations and Thiele approach based on the frustrated classical Heisenberg model on two-dimensional triangular lattice. A Hall-like motion is revealed in bulk system, similar to that in chiral magnets. More interestingly, our work demonstrates that the edge confinement in nanostripes of frustrated magnets can completely suppress the Hall motion and significantly accelerate the motion along the gradient direction.

The remainder of this manuscript is organized as follows: in Sec. II the model and the calculation method will be described. Sec. III is attributed to the results and discussion, and the conclusion is presented in Sec. IV.

## II. MODEL AND METHODS

Following the earlier work,[28] we consider the Hamiltonian

$$H = -J_1 \sum_{\langle i,j \rangle} \mathbf{S}_i \cdot \mathbf{S}_j + J_2 \sum_{\langle\langle i,j \rangle\rangle} \mathbf{S}_i \cdot \mathbf{S}_j - h \sum_i S_i^z - D \sum_i \left(S_i^z\right)^2 - \sum_i D_i' \left(S_i^z\right)^2, \quad (1)$$

where $\mathbf{S}_i$ is the classical Heisenberg spin with unit length on site $i$. The first term is the ferromagnetic NN interaction with $J_1 = 1$ (we use $J_1$ as the energy unit, for simplicity), and the second term is the antiferromagnetic NNN interaction with $J_2 = 0.5$, and the third term is the Zeeman coupling with a linear gradient field $h = h_0 + \mathbf{g} \cdot \mathbf{r}$ ($h_0 = 0.4$, $\mathbf{r}$ is the coordinate, and $\mathbf{g}$ is the gradient vector with a strength $g$) applied along the [001] direction,[28] and the fourth term is the bulk uniaxial anisotropy energy with $D = 0.2$, and the last term is the easy plane anisotropy energy of the edges with $D' = -2$. $D'$ is only considered at the edges for the nanostripes system, which may give rise to several types of edge states, as uncovered in earlier work.[28] However, it has been confirmed that the skyrmions/antiskyrmions in nanostripes move with the same speed when they are captured by one of these edge states. In this work, we mainly concern the gradient-driven motion of isolated skyrmions/antiskyrmion.

We study the spin dynamics at zero $T$ by numerically solving the LLG equation:

$$\frac{d\mathbf{S}_i}{dt} = -\gamma \mathbf{S}_i \times \mathbf{f}_i + \alpha \mathbf{S}_i \times \frac{d\mathbf{S}_i}{dt}, \quad (2)$$

with the local effective field $\mathbf{f}_i = -(\partial H/\partial \mathbf{S}_i)$. Here, $\gamma = 6$ is the gyromagnetic ratio, $\alpha$ is the Gilbert damping coefficient. We use the fourth-order Runge-Kutta method to solve the LLG equation. The initial spin configurations are obtained by solving the LLG equation at $g = 0$. Subsequently, the spin dynamics are investigated under gradient fields. Furthermore, the simulated results are further explained using the approach proposed by Thiele.[29] The displacement of the skyrmion/antiskyrmion is characterized by the position of its center ($X$, $Y$):

$$X = \frac{\int x(1-S^z)dxdy}{\int(1-S^z)dxdy}, \quad Y = \frac{\int y(1-S^z)dxdy}{\int(1-S^z)dxdy}. \tag{3}$$

Then, the velocity $\mathbf{v} = (v_x, v_y)$ is numerically calculated by

$$v_x = dX/dt, \quad v_y = dY/dt. \tag{4}$$

At last, $v_\perp$ and $v_\parallel$ are obtained through a simple coordinate transformation.

## III. RESULTS AND DISCUSSION

First, we investigate the spin configurations of possible isolated skyrmions and antiskyrmions with various vorticities and helicities obtained by LLG simulations of bulk system ($D' = 0$) at zero $g$. Specifically, four typical isolated skyrmions with the topological charge $Q = 1$ have been observed in our simulations, as depicted in Fig 1(a). The first two skyrmions are Néel-type ones with different helicities, and the remaining two skyrmions are Bloch-type ones. Furthermore, isolated antiskyrmoins are also possible in this system, and their spin configurations with $Q = -1$ are shown in Fig. 1(b).

After the relaxation of the spin configurations at $g = 0$, the magnetic field gradient is applied along the direction of $\theta = \pi/6$ ($\theta$ is the angle between the gradient vector and the positive $x$ axis, as shown in Fig. 2(a)) to study the dynamics of isolated skyrmions and antiskyrmions in bulk system. The LLG simulation is performed on a $28 \times 28$ triangular lattice with the periodic boundary condition applied along the $y'$ direction perpendicular to the gradient. Furthermore, we constrain the spin directions at the edges along the $x$ direction by $S^z = 1$ (red circles in Fig. 2(a)) to reduce the finite lattice size effect. Similar to that in chiral magnets, the skyrmion/antiskyrmion motion can be also driven by the magnetic field gradients in frustrated magnets. Fig. 2(b) and Fig. 2(c) give respectively the calculated $v_\parallel$ and $v_\perp$ as functions of $g$ at $\alpha = 0.04$. $v_\parallel$ of the skyrmion equals to that of the antiskyrmion, and both $v_\parallel$ and $v_\perp$ increase linearly with $g$. For a fixed $g$, the value of $v_\perp$ is nearly an order of magnitude larger than that of $v_\parallel$, clearly exhibiting a Hall-like motion of the skyrmions/antiskyrmions. It is noted that $v_\perp$ is caused by the gyromagnetic force which depends on the sign of the topological charge. Thus, along the $y'$ direction, the skyrmion and antiskyrmion move oppositely under the field gradient, the same as earlier report.[18] Moreover, $v_{//}$ is resulted from the dissipative force which is associated with the Gilbert damping. For example, the linear dependence of $v_{//}$ on the Gilbert damping constant $\alpha$ has been revealed in chiral magnets,[18] which still holds true for the frustrated magnets. The dependence of velocity on $\alpha$ at $g = 10^{-3}$ is depicted in Fig. 3, which clearly demonstrates that $v_{//}$ increases linearly and $v_\perp$ is almost invariant with the increase of $\alpha$.

Subsequently, the simulated results are qualitatively explained by Thiele equations:

$$\alpha \Gamma v_\| - G v_\perp = \gamma \frac{\partial H}{\partial X'}, \text{ and } G v_\| + \alpha \Gamma v_\perp = \gamma \frac{\partial H}{\partial Y'}, \quad (5)$$

with the skyrmion/antiskyrmoin center $(X', Y')$ in the $x'y'$ coordinate system. Here,

$$G = \int d^2 r \mathbf{S} \cdot \left( \frac{\partial \mathbf{S}}{\partial x} \times \frac{\partial \mathbf{S}}{\partial y} \right) = 4\pi Q, \text{ and } \Gamma = \int d^2 r \frac{\partial \mathbf{S}}{\partial x} \cdot \frac{\partial \mathbf{S}}{\partial x} = \int d^2 r \frac{\partial \mathbf{S}}{\partial y} \cdot \frac{\partial \mathbf{S}}{\partial y}. \quad (6)$$

For the frustrated bulk magnets with the magnetic field gradient applied along the $x'$ direction, there are

$$\frac{\partial H}{\partial X'} = g \sum_i \left( S_i^z - 1 \right) = gq, \text{ and } \frac{\partial H}{\partial Y'} = 0. \quad (7)$$

For $\alpha \ll 1$, $q$ is almost invariant and the velocities can be estimated from

$$v_\| \approx gq \frac{\alpha \gamma \Gamma}{G^2}, \ v_\perp \approx -gq \frac{\gamma}{G}. \quad (8)$$

Thus, a proportional relation between the velocity and field gradient is clearly demonstrated. Furthermore, $v_\perp$ is inversely proportional to $G$ and/or the topological charge $Q$, resulting in the fact that the skyrmion and antiskyrmion move along the $y'$ direction oppositely, as revealed in our simulations.

For current-induced motion of skyrmions, the lateral confinement can suppress the Hall motion and accelerate the motion along the current direction.[29-31] The confinement effects on the $h$-gradient driven skyrmion/antiskyrmion motion are also investigated in the nanostripes of frustrated magnets. For this case, the LLG simulation is performed on an 84 × 30 triangular-lattice with an open boundary condition along the $y$ direction. For convenience, the field gradient is applied along the $x$ direction. The easy plane anisotropy with $D' = -2$ is considered at the lateral edges, which gives rise to the edge state and in turn confines the skyrmions/antiskyrmions. Fig. 4(a) gives the time dependence of the $y$ coordinate of the skyrmion center for $\alpha = 0.04$ and $g = 10^{-3}$. It is clearly shown that the isolated skyrmion jumps into the channel at $Y = 17$ and then moves with a constant speed along the gradient (negative $x$, for this case) direction. Furthermore, the position of the channel changes only a little due to the small range of the gradient considered in this work, which never affects our main conclusions.

More interestingly, the skyrmion/antiskyrmion motion along the gradient direction can be significant accelerated by the lateral confinement, as shown in Fig. 4(b) which gives $v_{//}$ ($v_x$) as a function of $g$ at $\alpha = 0.04$. For a fixed $g$, $v_{//}$ of the nanostripes is almost two orders of magnitude larger than that of bulk system. When the skyrmion/antiskyrmion is captured by the edge state (under which $v_\perp = 0$), the equation (5) gives $v_{//} = gq\gamma/\alpha\Gamma$. It is shown that $v_{//}$ is inversely proportional to $\alpha$ in this confined geometry, and small $\alpha$ results in a high speed of motion of the skyrmion/antiskyrmion. The inversely proportional relation between $v_{//}$ and $\alpha$ has also been confirmed in our LLG simulations, as clearly shown in Fig. 4(c) which gives the simulated $v_{//}$ as a function of $1/\alpha$ at $g = 10^{-3}$.

At last, we study the effect of the reversed gradient **g** on the skyrmion/antiskyrmion motion, and its trail is recorded in Fig. 4(d). It is clearly shown that the reversed gradient moves the skyrmion/antiskyrmion out of the former channel near one lateral edge and drives it to the new channel near the other lateral edge. Subsequently, the skyrmion/antiskyrmion is captured by the new channel and moves reversely, resulting in the loop-like trail. As a result, our work suggests that one may modulate the moving channel by reversing the field gradient, which is meaningful for future applications such as in data erasing/restoring.

Anyway, it is suggested theoretically that the confined geometry in nanostripes of frustrated magnets could significantly speed up the field-driven motion of the isolated skyrmions/antiskyrmions, especially for system with small Gilbert damping constant. Furthermore, we would also like to point out that this acceleration is probably available in other confined materials such as chiral magnets, which deserves to be checked in future experiments.

## IV. CONCLUSION

In conclusion, we have studied the magnetic-field-gradient-driven motion of the isolated skyrmions and antiskyrmions in the frustrated triangular-lattice spin model using Landau-Lifshitz-Gilbert simulations and Thiele theory. The Hall-like motion is revealed in bulk system, similar to that in chiral magnets. More interestingly, it is suggested that the lateral confinement in the nanostripes of the frustrated system can suppress the Hall motion and significantly speed up the motion along the gradient direction. The acceleration of the motion is mainly determined by the Gilbert damping constant, which is helpful for finding potential materials for skyrmion-based spintronics.

**Acknowledgements**:

The work is supported by the National Key Projects for Basic Research of China (Grant No. 2015CB921202), and the National Key Research Programme of China (Grant No. 2016YFA0300101), and the Natural Science Foundation of China (Grant No. 51332007), and the Science and Technology Planning Project of Guangdong Province (Grant No. 2015B090927006). X. Lu also thanks for the support from the project for Guangdong Province Universities and Colleges Pearl River Scholar Funded Scheme (2016).

**FIGURE CAPTIONS**

Fig.1. Typical LLG snapshot of the spin configurations of skyrmions and antiskyrmions. (a) skyrmion structures and (b) antiskyrmion structures with different helicities.

Fig.2. (a) Effective model on the triangular lattice. (b) $v_{//}$ and (c) $v_\perp$ as functions of $g$ at $\alpha = 0.04$ in bulk system.

Fig.3. (a) $v_{//}$ and (b) $v_\perp$ as functions of $\alpha$ at $g = 10^{-3}$.

Fig.4. (a) Time dependence of $Y$ coordinate of the skyrmion center at $\alpha = 0.01$ and $g = 10^{-3}$. $v_{//}$ as a function of (b) $g$ at $\alpha = 0.04$, and (c) $\alpha$ at $g = 10^{-3}$ in the nanostripes of frustrated magnets. (d) The trail of skyrmion/antiskyrmion. The red line records the motion with gradient $\boldsymbol{g}$ along the positive $x$ direction, while the blue line records the motion with the reversed $\boldsymbol{g}$.

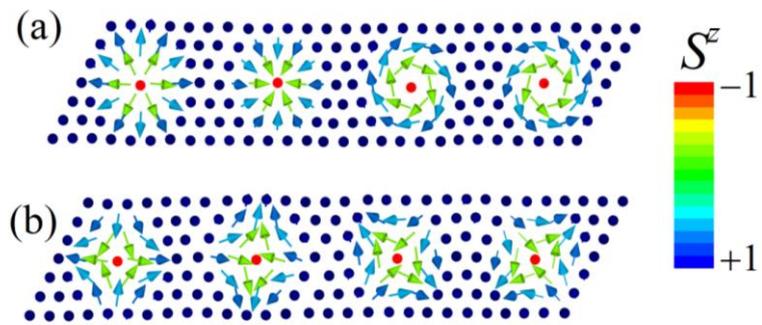

Fig.1. Typical LLG snapshot of the spin configurations of skyrmions and antiskyrmions. (a) skyrmion structures and (b) antiskyrmion structures with different helicities.

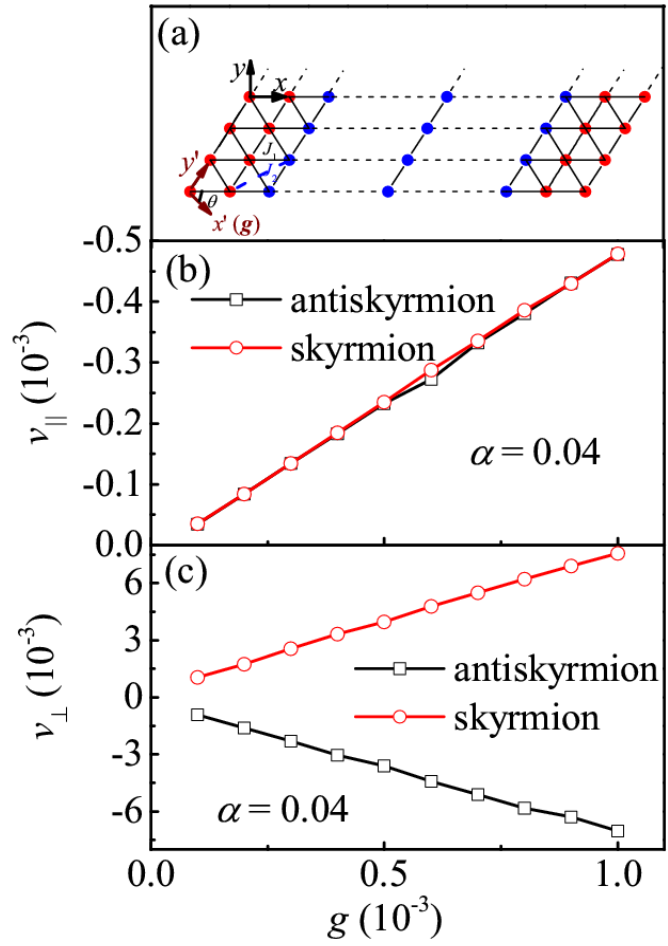

Fig.2. (a) Effective model on the triangular lattice. (b) $v_{//}$ and (c) $v_{\perp}$ as functions of $g$ at $\alpha = 0.04$ in bulk system.

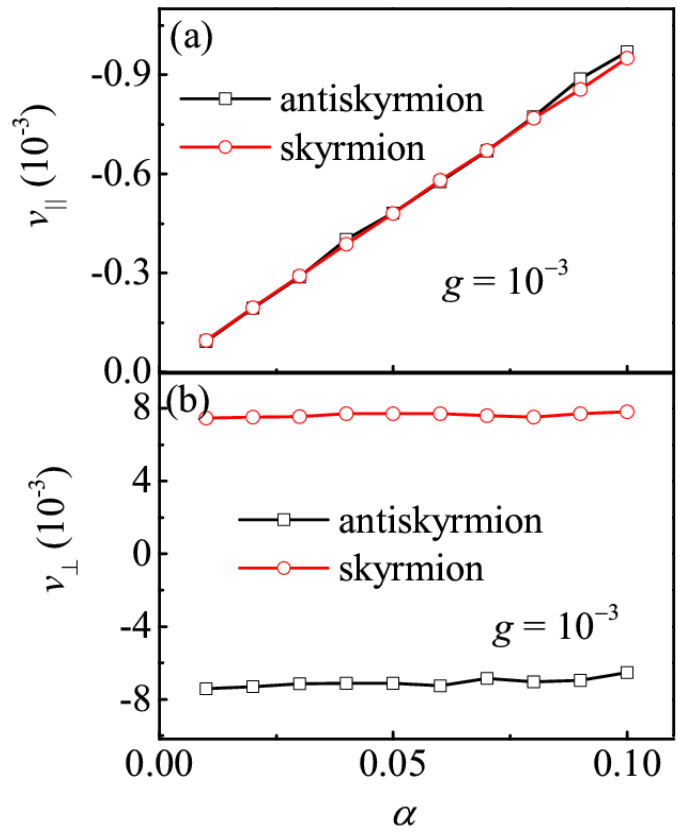

Fig.3. (a) $v_{//}$ and (b) $v_\perp$ as functions of $\alpha$ at $g = 10^{-3}$.

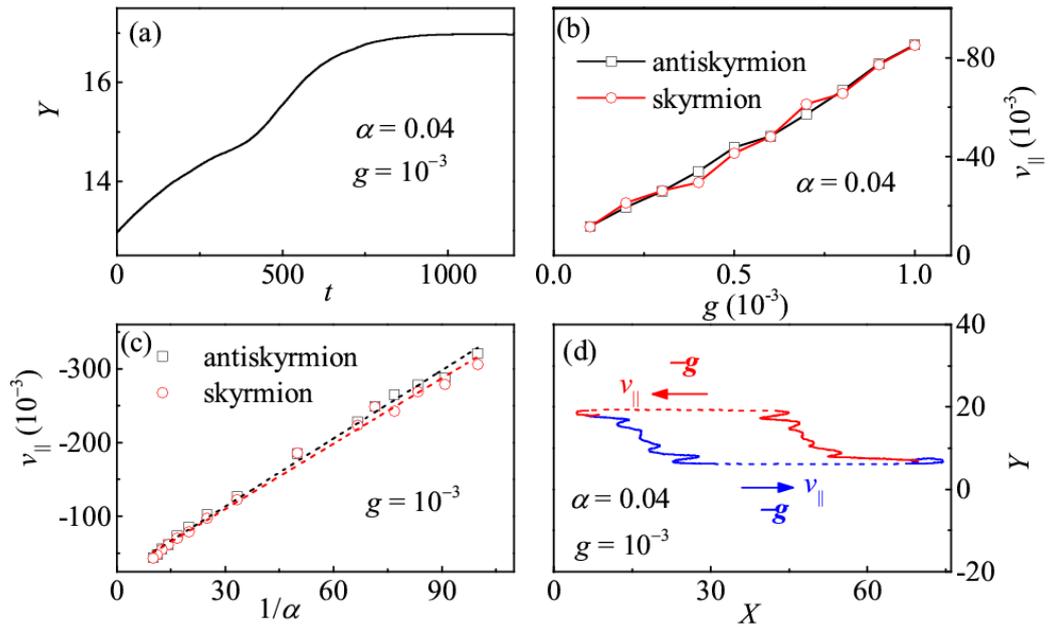

Fig.4. (a) Time dependence of $Y$ coordinate of the skyrmion center at $\alpha = 0.01$ and $g = 10^{-3}$. $v_{//}$ as a function of (b) $g$ at $\alpha = 0.04$, and (c) $\alpha$ at $g = 10^{-3}$ in the nanostripes of frustrated magnets. (d) The trail of skyrmion/antiskyrmion. The red line records the motion with gradient $g$ along the positive $x$ direction, while the blue line records the motion with the reversed $g$.